# Breaking of macroscopic centric symmetry in paraelectric phases of ferroelectric materials and implications for flexoelectricity

Alberto Biancoli[1], Chris M. Fancher[2], Jacob L. Jones[2] and Dragan Damjanovic[1,*]
[1]*Ceramics Laboratory, Swiss Federal Institute of Technology in Lausanne- EPFL, Lausanne, Switzerland*
[2]*Department of Materials Science and Engineering, North Carolina State University Raleigh, NC, USA*

**A centrosymmetric stress cannot induce a polar response in centric materials; piezoelectricity is, for example, possible only in non-centrosymmetric structures. An exception is meta-materials with shape asymmetry, which may be polarized by stress even when the material is centric. In this case the mechanism is flexoelectricity, which relates polarization to a strain gradient. The flexoelectric response scales inversely with size, thus a large effect is expected in nanoscale materials. Recent experiments in polycrystalline, centrosymmetric perovskites [e.g., (Ba,Sr)$TiO_3$] have indicated values of flexoelectric coefficients that are orders of magnitude higher than theoretically predicted, promising practical applications based on bulk materials. We show that materials with unexpectedly large flexoelectric response exhibit breaking of the macroscopic centric symmetry through inhomogeneity induced by the high temperature processing. The emerging electro-mechanical coupling is significant and may help to resolve the controversy surrounding the large apparent flexoelectric coefficients in this class of materials.**

The local or global breaking of symmetry and emergence of “forbidden” properties have been observed in many nominally centrosymmetric materials;[1-9] however, the origins of the symmetry breaking are not well understood. In the paraelectric phase of ferroelectric perovskites, local breaking of centric symmetry has been associated with polar precursors.[7,8,10-13] In some cases local polar nano-regions may align to give rise to the macroscopic polar symmetry. For instance, in amorphous films of $SrTiO_3$, partial alignment of distorted $TiO_6$ octahedra occurs under strain originating from the temperature gradient.[1] The breaking of the macroscopic centric symmetry has been reported in paraelectric phase of $BaTiO_3$ single crystals[14] and is often encountered in thin films.[15] The macroscopic symmetry breaking takes place in composite meta-materials with geometrically asymmetric structures.[16] The application of centrosymmetric pressure on such meta-materials generates strain gradient $\partial S_{ij}/\partial x_k$ and electrical polarization $P_l$ which are coupled through the flexoelectric effect by relationship $P_l = \mu_{ijkl}(\partial S_{ij}/\partial x_k)$ where $\mu_{ijkl}$ are components of the flexoelectric tensor and $x_k$ is direction of the strain gradient.[16,17] Unlike piezoelectricity, flexoelectricity is not limited by crystal symmetry and can appear in all sufficiently insulating materials.[16,18]

Theoretical models show that the flexoelectric response scales inversely with size and in nano-scale structures can be engineered to be as large as the piezoelectric effect.[19,20] Local strain gradients are common in complex materials and flexoelectricity was invoked to explain nano-scale inversion of polarization in $BaTiO_3$ films by pressure,[21] polarization rotation in $PbTiO_3$ films[22] and large properties at the morphotropic phase boundary in ferroelectrics.[23] Flexoelectricity was suggested as a mechanism for charge generation in biological systems.[24] A large flexoelectric response that is of a practical interest has been reported in studies of bulk composites based on paraelectric phases of polycrystalline ferroelectrics.[17,20,25,26] Ferroelectrics are of interest for flexoelectric devices because they exhibit large permittivity $\varepsilon$ and $\mu \propto \varepsilon$[18]. Surprisingly, the experimentally measured values of flexoelectric coefficients in ferroelectrics are up to two to four orders of magnitude larger[17,20,25,26] than theoretically predicted.[18,27,28] The origin of this large discrepancy is presently not understood.[28,29]

We present evidence of the breaking of macroscopic centric symmetry in paraelectric (centrosymmetric) phases of polycrystalline and single–crystal samples of perovskite $(Ba_{1-x}Sr_x)TiO_3$ solid solution. The symmetry breaking is observed in samples that are centrosymmetric by shape and in absence of macroscopic strain gradients caused by external forces. The ensuing polar symmetry gives rise to the electro-mechanical coupling which is of the same order of magnitude as reported for the flexoelectric effect in those materials[17] and may thus provide rationalization for the disparity between the experimental and theoretical values of flexoelectric coefficients.

Experiments were performed on $(Ba_{1-x}Sr_x)TiO_3$ ceramics (x=0, 0.025, 0.33, 0.40, 0.45, 0.50, 0.67, 0.90, 0.975, and 1) and crystals (x=0, 0.025, and 1). At ambient temperature (≈295 K) compositions with x<0.33 are ferroelectric, those with x≥0.33 are paraelectric, for x≈0.9 the materials exhibit relaxor character,[30] while $SrTiO_3$ (x=1) is an incipient ferroelectric. This choice of compositions thus covers a wide range of behaviors that may be relevant for flexoelectricity. $(Ba_{0.67}Sr_{0.33})TiO_3$ (BST67/33) with the Curie temperature $T_C$≈293 K was previously reported to exhibit unexpectedly high flexoelectric coefficients in the paraelectric phase,[17,31] while the experimental value for $\mu$ in $SrTiO_3$ agrees well with the theoretical predictions.[32] Ceramics were prepared by the solid-state method, $SrTiO_3$ crystals are standard substrates used for thin films deposition and $BaTiO_3$ crystals were solution grown (see Methods and section S0 of Supplementary material (SI)[33]). The charge-force relationship, herein referred to as the piezoelectric effect, was measured with a dynamic press and pyroelectric current was measured using a dynamic heating method (section S1 of SI[33]). Thermally stimulated currents (TSC) were measured on unpoled samples at a constant heating rate without applying a bias voltage. Local structural inhomogeneity in the form of microstrain was confirmed using Williamson-Hall analysis of X-ray diffraction (XRD) data from various sections of samples.[34] All techniques are explained in detail in Methods or SI.[33]

We start our study with a version of the classical flexoelectric experiment with truncated pyramid[17,20] by taking trapezoidal prisms cut from BST67/33 ceramic disks, Fig. 1. Because of the sample's shape asymmetry, a pressure applied along the symmetry axis (Fig. 1, **a**) generates strain gradient and charge via flexoelectric effect; the sign of the charge reverses when the prism is inverted by 180° (Fig. 1, panels **b** and **c**). Assuming that the charge is generated by the flexoelectric effect, we calculate $\mu$ to be ~$10^{-4}$ C/m, that is, of the same order of magnitude as reported previously[17,20] and some four orders of magnitude higher than predicted theoretically (2-20 nC/m).[18,27,28] The much larger experimental charge suggests that either the theoretical approach does not capture important details of the flexoelectric response, or that another type of electro-mechanical coupling operates simultaneously in the material that adds to and possibly dominates over the flexoelectric charge. It has been suggested that an extrinsic contribution to the flexoelectric effect could involve response of ferroelastic polar nano–regions to a strain gradient.[35]

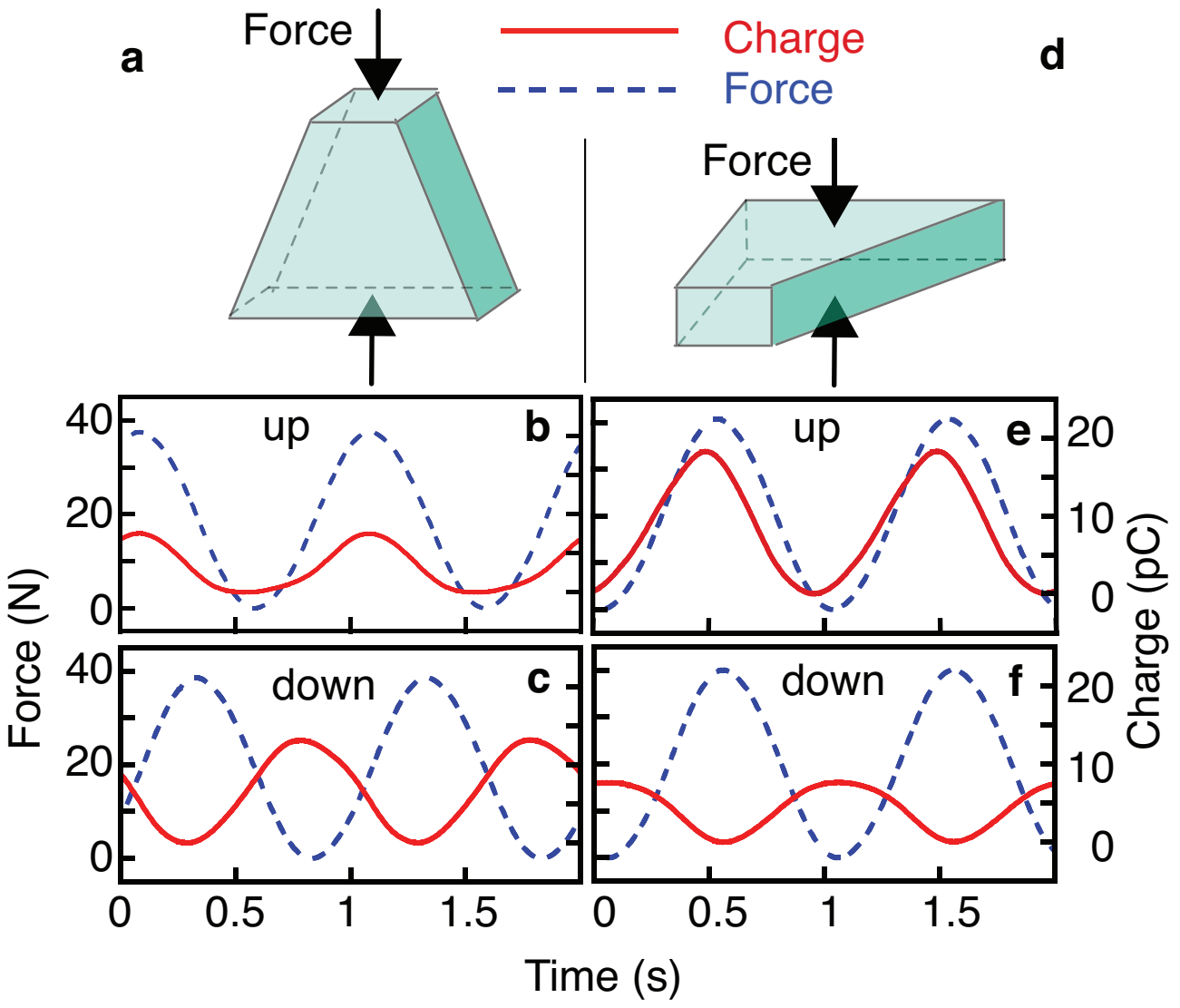

Fig. 1. **Flipped trapezoidal prism experiment.** "Up" designates an orientation of the sample and "down" the state rotated by 180°. While **a, b** and **c** may be interpreted as flexoelectric effect, **d**, **e** and **f** cannot. The sample is BST67/33 (height 5.30 mm, bases 7.96 $mm^2$ and 5.20 $mm^2$, thickness between 1.92 and 2.09 mm. Accuracy of dimension measurements: 0.01 mm). Static force ≈130 N.

To test whether the observed signal is indeed a result of the flexoelectric effect, we rotated the sample by 90° and applied pressure on the faces parallel to the symmetry axis (Fig. 1, **c**). Unexpectedly, the charge is again observed (Fig. 1, **e**) . Inversion of the rotated sample by 180° led to a signal of the opposite polarity (Fig. 1, **f**), indicating the absence of a center of symmetry along this direction. Because the application of a stress perpendicular to the symmetry axis should, ideally, not lead to a strain gradient and flexoelectric charge originating from the shape asymmetry, a question is posed on the origin of the charge in both directions.

It is straightforward to show that the cause of the piezoelectric response in rotated sample is neither the flexoelectric effect originating in geometrical imperfections nor a built-in strain gradient in our samples (see section S2 of SI[33]). To test whether the piezoelectric signal in Fig. 1 (**d, e, f**) is an artifact of the measuring method, which involves application of static and dynamic pressure (see Fig. S1.1 in SI[33]), we verified the absence of centric symmetry by inspecting whether the samples are pyroelectric. The temperature of samples was varied by approximately 1 K at a constant background temperature[33] and current measured. In this dynamic measurement mode, a centrosymmetric sample should not show a pyroelectric current; in contrast, polar materials (a subset of noncentrosymmetric materials) may exhibit a pyroelectric current, which is proportional to the temperature rate.[36] A clear pyroelectric current measured in a thin, disk-shaped sample of $(Ba_{0.60}Sr_{0.40})TiO_3$ (BST60/40) is shown in Fig. 2. With the $T_C$≈272 K, BST60/40 should not exhibit pyroelectric response at room temperature. Similar to the piezoelectric charge shown in Fig. 1, the pyroelectric current in BST60/40 reverses its direction with respect to the sense of temperature change when the sample is inverted by 180°, Fig. 2. These experiments demonstrate unequivocally that paraelectric BST60/40 samples are not only macroscopically noncentrosymmetric, but also polar. Moreover, the pyroelectric current in BST60/40 was measured up to 80 K above its Curie temperature (see section S3 of SI, Fig. S3.1[33]).

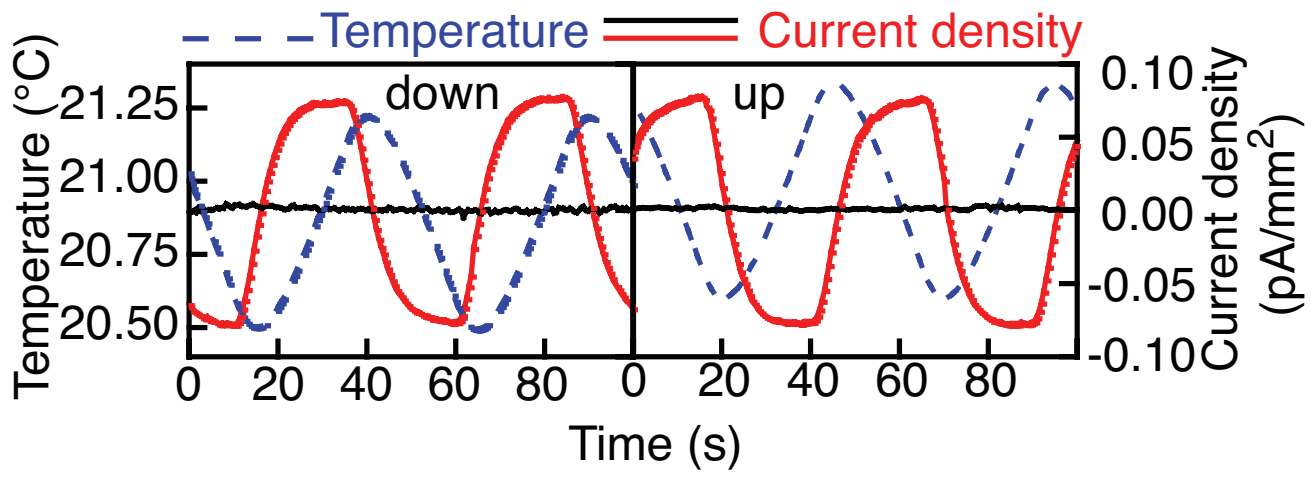

Fig. 2. **Pyroelectric current in BST60/40 and $SrTiO_3$.** Measurements were made at room temperature, ≈294 K, on BST60/40 and $SrTiO_3$ ceramics and on an $SrTiO_3$ single crystal, all with gold electrodes. The current in BST60/40 reverses direction with respect to the temperature change when the sample is inverted ("down" versus "up"). The flat lines near zero indicate currents in $SrTiO_3$ 001 oriented single crystal (left) and ceramic (right) measured under similar conditions as in BST60/40.

Similar piezoelectric charge and pyroelectric current were observed at room temperature in all samples, regardless of whether they possess paraelectric, ferroelectric or relaxor character; the only exceptions are $(Ba_{0.025}Sr_{0.975})O_3$ ceramics and $SrTiO_3$ ceramics and single crystals (see Fig. 2). The absence of signal in these samples could simply be related to the instruments' resolution; as shown below the polar nature of these samples was indeed confirmed by the TSC. Observation of piezoelectric and pyroelectric charge above $T_C$ in single crystals of $(Ba_{0.975}Sr_{0.025})TiO_3$ and $BaTiO_3$ shows that the symmetry breaking is not caused by the polycrystalline nature of ceramic samples (see Fig. S3.2 of SI[33]).

In nontextured and unpoled polycrystalline ferroelectrics, one expects centric macroscopic symmetry and zero piezoelectric and pyroelectric response even below $T_C$.[37] Polarity may be induced by the application of electric field, through a process called poling,[38] or by an initial application of an alternating electric signal.[39] However, unless explicitly stated, samples investigated in this work have not been subjected to even the smallest electric field. It is thus interesting that in unpoled ferroelectric ceramics the piezoelectric and pyroelectric responses were of the same order of magnitude below and above $T_C$ (see Fig. S3.3[33]).

The first indication that the origin of symmetry breaking in these samples may be, at least partly, extrinsic in nature is seen in the frequency dispersion of the piezoelectric and dielectric properties, shown in Fig. 3 for BST60/40. Both the permittivity and apparent piezoelectric coefficient exhibit a twofold increase as the frequency decreases from 100 Hz to 10 mHz. The theoretical flexoelectric effect calculated for the ionic lattice and the related piezoelectric coefficient, which is proportional to $\mu$ and $\varepsilon$,[17] should not be frequency dependent in this frequency range. Low-frequency dispersion of permittivity in perovskite materials is usually attributed to space charges. While surface space charges may lead to apparent symmetry breaking in the paraelectric phase of ferroelectrics,[6] surface effects can be eliminated as a dominant reason for the symmetry breaking in materials examined in this study (see section S4 of SI[33]). We next give evidence of built-in polarization in the samples and its interaction with space charges.

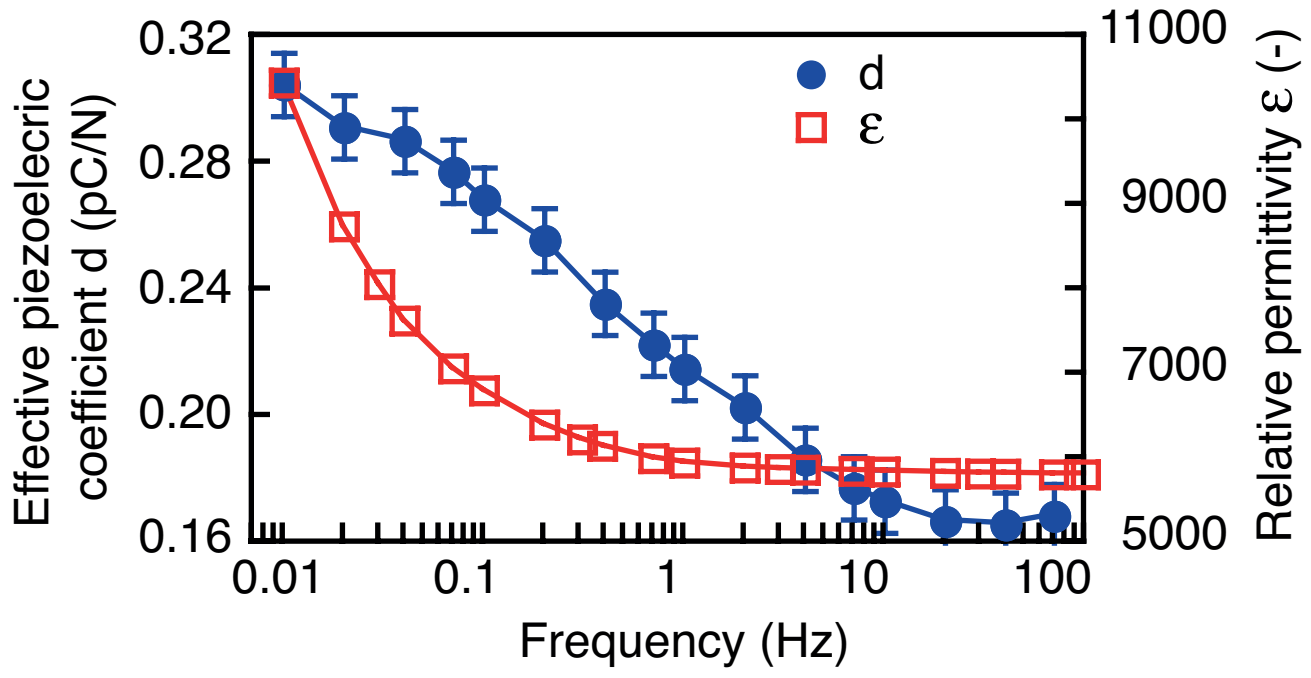

Fig. 3. **The frequency dependence of properties.** The apparent piezoelectric coefficient $d$ is defined as $Q_{max}/F_{max}$ ratio, where $Q_{max}$ is the charge amplitude and $F_{max}$ is the force amplitude. The relative permittivity ε was calculated from the capacitance determined as $Q_{max}/V_{max}$ ratio, where $Q_{max}$ is the charge amplitude and $V_{max}$ is the voltage amplitude. Measurements were made at 294 K on a BST60/40 sample.

All sintered ceramic samples exhibit polarization oriented in the same direction with respect to their position in the furnace (see section S5 of SI[33]). The polarization direction was determined from the phase of the piezoelectric and pyroelectric responses, and is also indicated by the sign of the TSC. The TSC were collected at zero electric field during heating (followed by cooling) of samples at a constant rate of 2.5 K/min from the room temperature to 820 K and back. Note that, as opposed to usual TSC experiments where samples are first subjected to an electric field to form an electret,[40] here we measured unpoled samples. In general, the origins of TSC include the field-induced polarization in electrets, built–in or spontaneous polarization, small voltage bias on the picoammeter, and the thermoelectric effect.[41]

The two latter mechanisms are not sensitive to the orientation of the sample. Since the current in our paraelectric samples was sensitive to sample orientation (Fig. 4 **a**), and samples exhibited pyroelectric current upon cooling to room temperature, the samples must exhibit a stable built–in polarization. It is this built–in polarization that defines the direction of the current. In all examined ceramics the TSC spectra exhibited two peaks in the range ~470 K to 670 K. Both peaks were absent on cooling (see Fig. S6.1 in SI[33]) showing that they are dominated by the release of trapped charges and not by the built–in polarization itself. The absence of peaks on cooling indicates that the peaks do not arise from phase transitions within polar nano–regions, as was proposed for relaxor ferroelectrics.[42] Immediately upon cooling a sample, the pyroelectric current at ambient temperature exhibited strong drift and its amplitude varied with time; we interpret this as settling of space charges that were excited during the heat treatment. After settling, which may last a few hours or days, the response of the material was usually increased with respect to the response observed immediately upon cooling. The space charge polarization thus reinforces the built–in polarization.

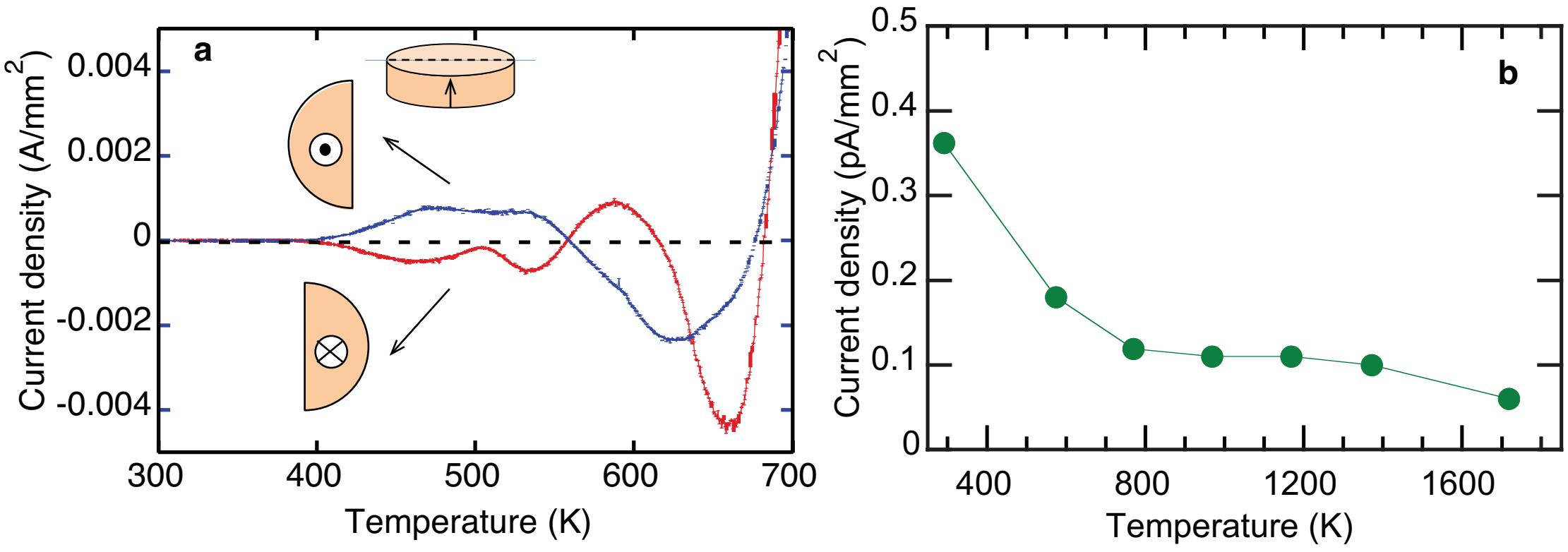

Fig. 4. **Thermally stimulated and pyroelectric currents.** (**a**) An as-sintered, electroded BST60/40 sample was cut into two halves and one half was turned by 180° with respect to the other. TSC was measured on each half during the heating (shown) and cooling (not shown here, see Fig. S6.1). The different signs of the TSC indicate built–in polarization in the samples. The two peaks between 470 K and 570 K indicate release of space charges. The peaks are absent on cooling. The large current increase above 670 K, which does not depend on sample orientation, could be due to thermo-electric currents or due to low resistance of the sample at high temperature and small bias voltage across the picoammeter. The large minima above 570 K are a typical result of competing currents of different signs (Ref. 40). (**b**) The amplitude of pyroelectric current of a BST60/40 sample measured at room temperature after annealing at different temperatures for 1 hour. The sample was heated and cooled under open circuit conditions, with or without top electrodes.

Further evidence for the presence of a stable built–in polarization and volatile space charge contribution to the polarization of samples was obtained by annealing the samples at high temperatures (up to 1770 K), Fig. 4 **b**. Annealing above 770 K led to a large decrease in pyroelectric current at room temperature, which then remained stable even after annealing at 1770 K. The built–in polarization is robust and survives such extreme temperatures, while the component of the total polarization that is built of mobile space charges is lost during the high temperature annealing. In samples that were poled to form an electret (see section S6 of SI[33]), the polar response was either increased multifold or reversed with respect to the built–in polarization, depending on the direction of the poling field. However, the space charges that were rearranged by the poling field decayed within few days, restoring the polarization before poling (Fig. S6.2 in SI[33]).

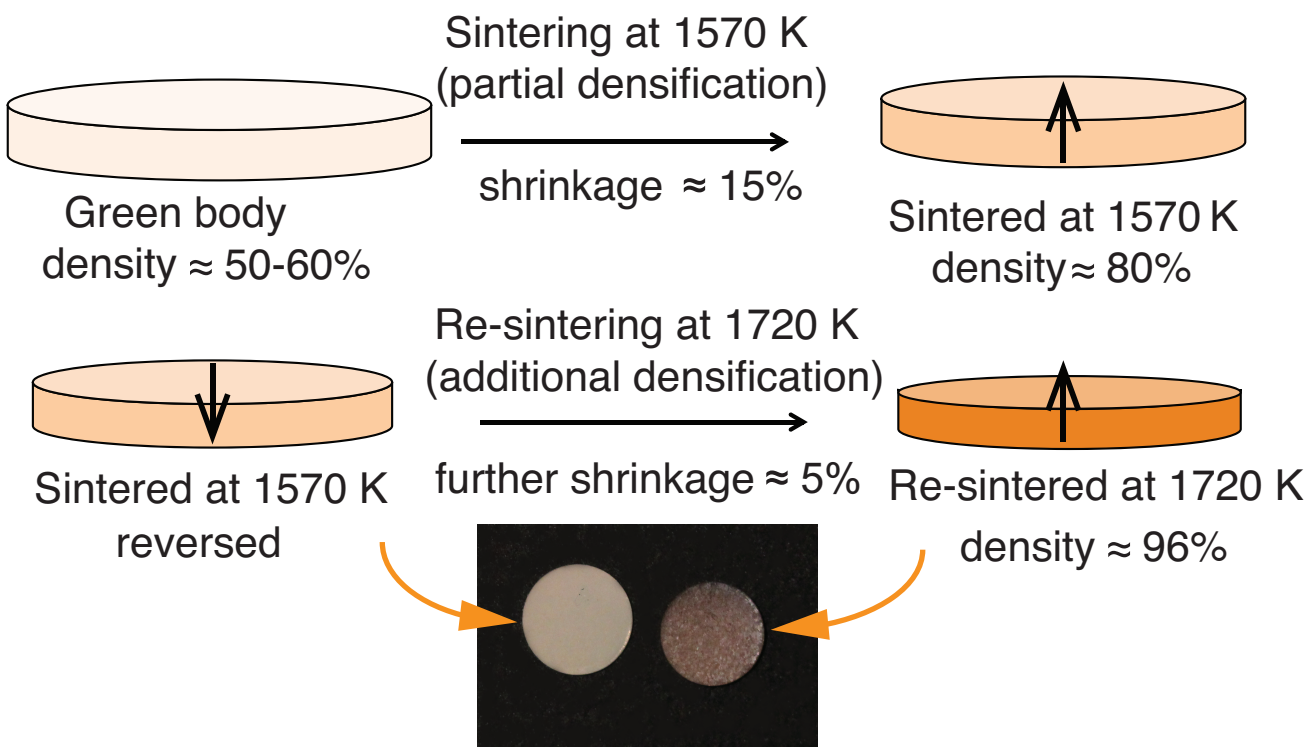

**Fig. 5. The built-in polarization and its direction**. The polarization develops while the sample densifies during sintering, adopting the preferred direction with respect to the samples' orientation in the furnace (see section S5, SI[33]). If a sample is partially densified (density 80% of the theoretical) at 1570 K, the polarization develops along a preferred direction (see the disc on the upper right corner marked with upward arrow). If the sample is then cooled and reversed (the disc on the lower left part with downward arrow) and densification continued at 1720 K, the polarization develops again along the preferred direction (disc on the lower right part with upward arrow). This indicates significant redistribution of inhomogeneities linked to the densification process during the sintering. The densification effect on the sample size and color is clearly seen in the photographs of the samples sintered at 1570 K and at 1720 K, shown at the bottom of the figure.

We next discuss possible origins of the built-in polarization and show that it is most likely formed during the densification step. We are guided by the consistent observation that the direction of built–in polarization is determined by the orientation of the sample in the furnace during sintering (see S5 in SI[33]). The sintering is the process during which grains of the pressed ceramic powder bond together and the sample densifies by diffusion. The transport of material and mobility of ions are maximal during sintering, as seen in the typical change of the sample's dimensions by ≈20%. We demonstrate with the experiment depicted in Fig. 5 that the polarization direction in the samples is set–in during densification. Firing of a sample at 1570 K leads to partial densification (density up to 80% of the theoretical value) accompanied by a built–in polarization with a direction that is set by the orientation of the sample in the furnace. If the sample is then flipped over and the firing continued at a higher temperature (1720 K) the sample reaches a density of ≈96% of the theoretical, and the polarization inverts to conform to the preferred direction. We infer from this result that during the densification from 80% to 96% of the theoretical density, additional mass transport takes place, which redistributes charged defects and reconstructs the preferred polarization direction. Importantly, once nearly full densification is reached, the polarization direction is stable and can no longer be changed even if the sample is inverted and heated well above the sintering temperature. We propose that, in single crystals, the redistribution of charges and polarization imprint in the paraelectric phase may occur during the crystal growth.

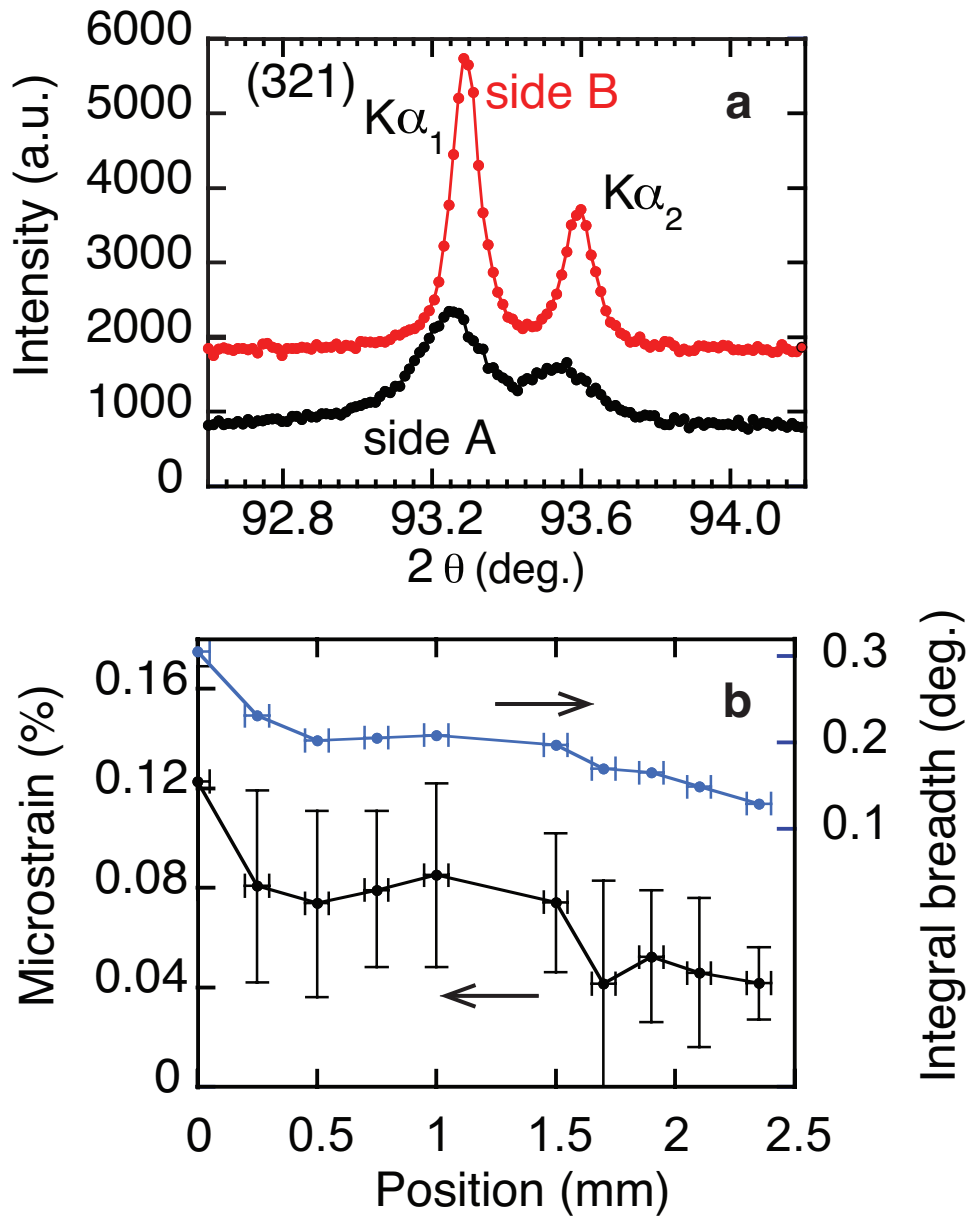

Fig. 6. **XRD analysis of a sintered BST60/40 sample.** (**a**) The XRD patterns of the 321 reflections from the top (side B) and bottom (side A) surfaces of a sample demonstrate a clear difference in peak profile. (**b**) Williamson-Hall analysis of XRD patterns measured after sequentially removing ~200 μm determines the presence of a gradient in microstrain through the sample thickness. The vertical error bars represent 95% confidence interval, and horizontal indicate accuracy in determination of the thickness of removed layers.

To inspect for the presence of inhomogeneities that must be associated with charge separation and ensuing built-in polarization, we analyzed samples chemically and structurally through the bulk of material. Energy dispersive X-ray spectroscopy (EDX) measurements, although suggesting a gradient of Ba/Sr ratio through the cross section of a BST60/40 sample, were inconclusive because of the poor resolution of this technique. X-ray diffraction (XRD) patterns were measured from both sides of several sintered samples, and then on planes at various distances between and parallel to the original surfaces. The XRD patterns at these internal positions were measured after sequentially removing ~200 μm of the material from both sides, Fig. 6. A clear difference in peak profile shapes (Fig 6 **a**) is observed on the two sintered surfaces and changes consistently through the thickness of the sample. A Williamson-Hall analysis of the integral breadths of the XRD patterns was used to characterize the crystallite size and microstrain (i.e., distribution of lattice constants) through the thickness of the samples (Fig 6 **b**). For the same composition, these measurements were consistent across samples of various shapes and sizes. Importantly, the microstrain was smaller in $SrTiO_3$, whereas $BaTiO_3$-rich compositions exhibited more microstrain overall (suggesting microscale inhomogeneity in $BaTiO_3$-rich compositions) and a change in microstrain through the sample thickness (suggesting mesoscale inhomogeneity through the sample). We thus show that the polarization imprint takes place during densification and is accompanied by the inhomogeneous lattice strain detected throughout the bulk of the samples. Similar strain build-up may happen during the growth of single crystals and thin films, although its origin may be different than in ceramics.

There are several ways in which inhomogeneous strain may interact with charges and lead to macroscopic built-in polarization. Charged oxygen vacancies, $V_O^{+2}$, are the most common ionic defect in perovskite materials, and are known to distribute inhomogeneously if a driving electrical field is provided.[43,44] In the absence of electric field, the driving force for the inhomogeneous distribution could be strain. *Ab-initio* calculations have shown that the formation energy of oxygen vacancies in perovskites may be lower under tensile strain;[45] alternatively, oxygen vacancies increase the volume of material (i.e., lead to strain), the concept known as chemical expansivity.[46] In materials such as those investigated here where non-negligible gradient in microstrain is observed throughout the sample, one can therefore expect a gradient in concentration of oxygen vacancies. However, the intentional introduction of an oxygen vacancy gradient in $SrTiO_3$ ceramics by annealing samples with one side in contact with a graphite block to promote reduction did not lead to a detectable pyroelectric response.

Because the built-in polarization sets in at the high sintering temperature, the Schottky defects (cation-anion vacancy pairs) cannot be excluded *a priori*, even though their energy of formation is higher than that of anionic defects.[47,48] An origin of the A-site vacancies could be precipitation of a Ba-rich phase. Indeed, we have detected small amounts of $Ba_2TiO_4$ phase in Ba-rich compositions by the EDX. This could explain why pyroelectric current is not observed in $SrTiO_3$ or $(Sr_{0.975}Ba_{0.025})TiO_3$ at room temperature. However, since the sign of the TSC in $SrTiO_3$ ceramics and single crystals is sensitive to the sample orientation (see Fig. S6.3 in SI[33]), the symmetry is obviously broken even in these samples, but the built-in polarization is either too small or too rigid to be changed by the weak temperature and pressure variation used for piezoelectric and pyroelectric measurements at ambient temperature.

Another candidate for the origin of the built-in macroscopic polarization is the strain-driven alignment of the polar nano–entities that are formed on the lattice level in the paraelectric phase, such as polar clusters[7,8] or polar nano-regions.[1,10] If elastically active, these polar entities could freeze along a preferred orientation in presence of strain gradients that develop during the sintering process and crystal growth, or by epitaxial strains, as shown by first-principle calculations in $Ba(Zr,Ti)O_3$ thin films.[13] Because of different sizes of Ba and Sr cations, the polar regions in Sr– and Ba–rich compositions are expected to be different; for example, Levin *et al.*[49] showed that off–centering of Ti cation in cubic $(Ba,Sr)TiO_3$ decreases as Ba/Sr ratio decreases. However, unless exhibiting a low geometrical symmetry (e.g., symmetry of the cone instead of usually assumed ellipsoidal symmetry), the strain–oriented polar entities are not expected to adopt the same sense, a necessary condition for the observed breaking of the macroscopic symmetry.

The volatile component of the polarization and low frequency effects reported here are likely due to the electronic defects, or a fraction of ionic defects that is not permanently trapped by strain but, for example, at grain boundaries.[44]

Finally, we address the origin of the asymmetry that leads to inhomogeneous strain. At the sintering and crystal–growth conditions where ion mobility is high, slight chemical and temperature gradients across the sample (see section S2 in SI[33]), or even gravity[50] could contribute to the anisotropic macroscopic strain fields that drive inhomogeneous distribution of charges or polar entities. The exact mechanisms are not clear, but appear to be pervasive in processing of ceramics, crystals and thin films. We have observed the same effects described here in other materials and in $(Ba,Sr)TiO_3$ samples prepared in other laboratories by different techniques.

In conclusion, our work reveals macroscopic symmetry breaking in paraelectric phases of ferroelectirc perovskite materials that can account for the unexpectedly large values of the apparent flexoelectric coefficients. The strain-driven charge redistribution appears to be characteristic for commonly used high–temperature processing methods which are accompanied with significant material transport. We show that the macroscopic breaking of centric symmetry in ferroelectrics is as common as the local symmetry breaking, but much less understood. The results further imply that the "self–polarization" reported in ferroelectric thin films may have similar strain–driven origin as described here, rather than being of purely electrical character as often assumed. The results should thus inspire theoretical and experimental work on the nature of interactions between strain and ionic defects and polar nano-regions in ferroelectrics and oxide materials.

**Methods:**
$(Ba_{1-x}Sr_x)TiO_3$ ceramics were prepared from $BaTiO_3$ and $SrTiO_3$ powders (see SI[33]). Powders were mixed 24 hours in isopropanol with $ZrO_2$ balls. Stoichiometric mixtures were calcined from 2 to 5 hours at 1150 °C. Some powders were mixed again for 24 hours and the calcination was repeated. A binder (polyvinyl alcohol) was added to the calcined powders (0.5% by weight of powders). The mixture was homogenized by mixing for 8 h in isopropanol with $ZrO_2$ balls or in a dry condition with a mortar and pestle. Powders were uniaxially pressed at 100 MPa to150 MPa in a cylindrical mold in shape of disks and sintered on different support materials, typically a Pt foil, at temperatures ranging from 1380 °C to 1450 °C for 1.5 to 4 hours. $SrTiO_3$ ceramics were prepared by adding polyvinyl alcohol as binder to $SrTiO_3$ powder, and the two were mixed in a mortar with a pestle. The powder was pressed into ≈0.5 mm thick disks with diameter ≈5.50 mm at 100MPa and sintered at 1500°C for 4 hours on a Pt foil. Single crystals of $SrTiO_3$ are standard substrates for deposition of thin films with <001> direction perpendicular to the plane of the plate. Density of the ceramic samples was ≈95-96 % of the theoretical. The grain size was in the range 5-10 μm for $x \geq 0.33$ and was ~50 μm for $x<0.33$ and did not itself qualitatively affect the response of the samples. The structure and phase content were inspected with X-ray diffraction (Bruker D8 Discover) on the surface of sintered pellets showing an expected perovskite phase without secondary phases (to the extent of XRD resolution). A PANalytical Empyrean was used for determination of microstrain. Stoichiometry was also verified by measuring the Curie temperature, as shown in Fig. S3.1 of SI. For pyroelectric and TSC measurements, Au electrodes were deposited by sputtering and currents measured on as-sintered or polished samples, with thickness of 0.3-0.5 mm. Samples for electro-mechanical measurements were used as–sintered or polished and without or with electrodes. A homemade dynamic press was used for charge-force measurements and its schematic is shown in Fig. S1 of SI.[33] For piezoelectric measurements, the accuracy is about 0.1 pC/N with precision of 0.01 pC/N. Pyroelectric measurements were conducted with a homemade set-up. The temperature of the sample was varied with a Peltier element at frequency range between 5 mHz and 50 mHz, and pyroelectric current determined by measuring voltage through a resistor with sub pico-Amper resolution. See section S1 of SI[33] for details. Dielectric dispersion measurements were made by applying 1 Vrms signal on an electroded sample. The generated charge was measured with a Kistler 5011 charge amplifier and SR830 DSP Lock-in amplifier. Capacitance could be determined with accuracy of 1-5 pF with resolution better than 1 pF. Thermally simulated currents were measured with a constant rate of 2.5 K/min, using Keithly 486 picoammeter. The input voltage burden of the instrument is 200 μV.

**Acknowledgment:**
This work was supported by the Swiss National Science Foundation through NRP62 program "Smart materials" (Project No. 406240 -126091).

**Author contributions:**
A.B. prepared all materials and performed most of the experiments. D.D. conceived the idea of testing the symmetry breaking and its effect on the apparent flexoelectric response and performed some of the experiments. A.B. and D.D. analyzed and interpreted the electrical data. C.M.F. performed XRD measurements and analyzed the data under supervision of J.L.J. D.D. wrote the article and all authors contributed and commented to the text.

**Competing Interests:** The authors declare that they have no competing financial interests.

**References:**
1. Frenkel, A. *et al.* Origin of polarity in amorphous $SrTiO_3$. *Physical Review Letters* **99**, 215502 (2007).
2. Scott, J. F. Lattice perturbations in $CaWO_4$ and $CaMoO_4$. *The Journal of Chemical Physics* **48**, 874-876 (1968).
3. Fox, G. R., Yamamoto, J. K., Miller, D. V., Cross, L. E. & Kurtz, S. K. Thermal hysteresis of optical second harmonic in paralelectric $BaTiO_3$. *Materials Letters* **9**, 284-288 (1990).

4. Wieczorek, K. *et al.* Electrostrictive and piezoelectric effect in $BaTiO_3$ and $PbZrO_3$. *Ferroelectrics* **336**, 61-67 (2006).
5. Beige, H., Birkholz, C., Ciesla, E. & Schmidt, G. Elastic and electromechanical properties of some crystals near structural phase transitions. *Physica Status Solidi B-Basic Research* **76**, K47-K50 (1976).
6. Lines, M. E. & Glass, A. M. *Principles and Applications of Ferroelectrics and Related Materials*. (Clarendon, Oxford, 1979).
7. Kleemann, W., Schafer, F. J. & Fontana, M. D. Crystal optical studies of spontaneous and precursor polarization in $KNbO_3$. *Phys. Rev. B* **30**, 1148–1154 (1984).
8. Aktas, O., Carpenter, M. A. & Salje, E. K. H. Polar precursor ordering in $BaTiO_3$ detected by resonant piezoelectric spectroscopy. *Applied Physics Letters* **103**, 142902 (2013).
9. Morozovska, A. N., Eliseev, E. A., Kalinin, S. V., Qing Chen, L. & Gopalan, V. Surface polar states and pyroelectricity in ferroelastics induced by flexo-roto field. *Applied Physics Letters* **100**, 142902 (2012).
10. Burns, G. & Dacol, F. Crystalline ferroelectrics with glassy polarization behavior. *Physical Review B* **28**, 2527-2530 (1983).
11. Bussmann-Holder, A., Beige, H. & Völkel, G. Precursor effects, broken local symmetry, and coexistence of order-disorder and displacive dynamics in perovskite ferroelectrics. *Physical Review B* **79**, 184111 (2009).
12. Kuroiwa, Y. *et al.* High-energy SR powder diffraction evidence of multisite disorder of Pb atom in cubic phase of $PbZr_{1-x}Ti_xO_3$. *Japanese Journal of Applied Physics* **44**, 7151-7155 (2005).
13. Prosandeev, S., Wang, D. & Bellaiche, L. Properties of epitaxial films made of relaxor ferroelectrics. *Physical Review Letters* **111**, 247602 (2013).
14. Bursian, E. V. & Zaikovski, O. I. Changes in the curvaure of a ferroelectric film due to polarization. *Soviet Physics - Solid State* **10**, 1121-1124 (1968).
15. Luo, Y. *et al.* Upward ferroelectric self-poling in (001) oriented $PbZr_{0.2}Ti_{0.8}O_3$ epitaxial films with compressive strain. *AIP Advances* **3**, 122101 (2013).
16. Fousek, J., Cross, L. E. & Litvin, D. B. Possible piezoelectric composites based on the flexoelectric effect. *Materials Letters* **39**, 287-291 (1999).
17. Cross, L. E. Flexoelectric effects: Charge separation in insulating solids subjected to elastic strain gradients. *Journal of Materials Science* **41**, 53-63 (2006).
18. Tagantsev, A. K. Piezoelectricity and flexoelectricity in crystalline dielectrics. *Phys. Rev. B* **34**, 5883-5889 (1986).
19. Majdoub, M. S., Sharma, P. & Cagin, T. Enhanced size-dependent piezoelectricity and elasticity in nanostructures due to the flexoelectric effect. *Physical Review B* **77**, 125424 (2008).
20. Fu, J. Y., Zhu, W. Y., Li, N., Smith, N. B. & Cross, L. E. Gradient scaling phenomenon in microsize flexoelectric piezoelectric composites. *Applied Physics Letters* **91**, 182910 (2007).

21. Lu, H. *et al.* Mechanical writing of ferroelectric polarization. *Science* **336**, 59-61 (2012).
22. Catalan, G. *et al.* Flexoelectric rotation of polarization in ferroelectric thin films. *Nature Materials* **10**, 963-967 (2011).
23. Borisevich, A. Y. *et al.* Atomic-scale evolution of modulated phases at the ferroelectric-antiferroelectric morphotropic phase boundary controlled by flexoelectric interaction. *Nature Communications* **3**, 775 (2012).
24. Lakes, R. The role of gradient effects in the piezoelectricity of bone. *IEEE Transactions on Biomedical Engineering* **27**, 282-283 (1980).
25. Ma, W. H. & Cross, L. E. Flexoelectric effect in ceramic lead zirconate titanate. *Applied Physics Letters* **86**, 072905 (2005).
26. Chu, B., Zhu, W., Li, N. & Cross, L. E. Flexure mode flexoelectric piezoelectric composites. *J. Appl. Phys.* **106**, 104109 (2009).
27. Ponomareva, I., Tagantsev, A. K. & Bellaiche, L. Finite-temperature flexoelectricity in ferroelectric thin films from first principles. *Physical Review B* **85**, 104101 (2012).
28. Hong, J. & Vanderbilt, D. First-principles theory and calculation of flexoelectricity. *Physical Review B* **88**, 174107 (2013).
29. Yudin, P. V., Ahluwalia, R. & Tagantsev, A. K. Upper bounds for flexoelectric coefficients in ferroelectrics. *Applied Physics Letters* **104**, 082913 (2014).
30. Zhou, L., Vilarinho, P. M. & Baptista, J. L. Dependence of the structural and dielectric properties of $Ba_{1-x}Sr_xTiO_3$ ceramic solid solutions on raw material processing. *J. Europ. Ceram. Soc.* **19**, 2015-2020 (1999).
31. Ma, W. H. & Cross, L. E. Flexoelectric polarization of barium strontium titanate in the paraelectric state. *Applied Physics Letters* **81**, 3440-3442 (2002).
32. Zubko, P., Catalan, G., Buckley, A., Welche, R. L. & Scott, J. F. Strain-gradient-induced polarization in $SrTiO_3$ single crystals. *Phys. Rev. Lett.* **99**, 167601 (2007).
33. Supplementary information.
34. Williamson, G. K. & Hall, W. H. X-ray line broadening from filed aluminium and wolfram. *Acta Metallurgica* **1**, 22-31 (1953).
35. Narvaez, J. & Catalan, G. Origin of the enhanced flexoelectricity of relaxor ferroelectrics. *Applied Physics Letters* **104**, 162903 (2014).
36 Chynoweth, A. G. Dynamic method for measuring the pyroelectric effect with special refernce to barium titanate. *J. Appl. Phys* **27**, 78-84 (1956).
37. Newnham, R. E. *Properties of Materials: Anisotropy, Symmetry, Structure*. (Oxford University, 2005).
38. Jaffe, B., Cook, W. R. & Jaffe, H. *Piezoelectric Ceramics*. (Academic, New York, 1971).
39. Dec, J. *et al.* Probing polar nanoregions in $Sr_{0.61}Ba_{0.39}Nb_2O_6$ via second-harmonic dielectric response. *Phys. Rev. B* **68**, 092105 (2003).
40. van Turnhout, J. *Thermally Stimulated Discharge of Polymer Electrets: A Study on Nonisothermal Dielectric Relaxation Phenomena*. (Elsevier, Amsterdam, 1975).

41. Lau, W. S., Chong, T. C., Tan, L. S., Goo, C. H. & Kian, S. G. The Characterization of traps in semi-insulating gallium arsenide buffer layers grown at low temperature by molecular beam epitaxy with an improved zero-bias thermally stimurated current technique. *Jpn. J. Appl. Phys.* **30**, L1843-L1846 (1991).
42. Roth, M., Mojaev, E., Dul'kin, E., Gemeiner, P. & Dkhil, B. Phase transition at a nanometer scale detected by acoustic emission within the cubic phase $Pb(Zn_{1/3}Nb_{2/3})O_3$-$xPbTiO_3$ relaxor ferroelectrics. *Physical Review Letters* **98**, 265701 (2007).
43. Cao, Y., Shen, J., Randall, C. A. & Chen, L. Q. Phase-field modeling of switchable diode-like current-voltage characteristics in ferroelectric $BaTiO_3$. *Applied Physics Letters* **104**, 182905 (2014).
44. Liu, W. & Randall, C. A. Thermally stimulated relaxation in Fe-doped $SrTiO_3$ Systems: II. Degradation of $SrTiO_3$ dielectrics. *Journal of the American Ceramic Society* **91**, 3251-3257 (2008).
45. Aschauer, U., Pfenninger, R., Selbach, S. M., Grande, T. & Spaldin, N. A. Strain-controlled oxygen vacancy formation and ordering in $CaMnO_3$. *Physical Review B* **88**, 054111 (2013).
46. Adler, S. B. Chemical Expansivity of Electrochemical Ceramics. *J. Am. Ceram. Soc.* **84**, 2117-2119 (2001).
47. Erhart, P. & Albe, K. Thermodynamics of mono- and di-vacancies in barium titanate. *Journal of Applied Physics* **102**, 084111 (2007).
48. Lewis, G. V. & Catlow, C. R. A. Defect studies of doped and undoped barium titanate using computer simulation techniques. *J. Phys. Chem. Sol.* **47**, 89-97 (1986).
49. Levin, I., Krayzman, V. & Woicik, J. C. Local structure in perovskite $(Ba,Sr)TiO_3$: Reverse Monte Carlo refinements from multiple measurement techniques. *Phys. Rev. B* **89**, 024106 (2014).
50. Lenel, F. V., Hausner, H. H., Roman, O. V. & Ansell, G. S. The influence of gravity in sintering. *Powder Metall Met Ceram* **2**, 379-384 (1963).

# Breaking of macroscopic centric symmetry in paraelectric phases of ferroelectric materials: implications to flexoelectricity

Alberto Biancoli[1], Chris M. Fancher[2], Jacob L. Jones[2] and Dragan Damjanovic[1,*]

[1]*Ceramics Laboratory, Swiss Federal Institute of Technology in Lausanne- EPFL, Lausanne, Switzerland*
[2]*Department of Materials Science and Engineering, North Carolina State University, Raleigh, NC, USA*

*dragan.damjanovic@epfl.ch

**S0 details on ceramics preparation and single crystals.** The following powders were used for preparation of samples: $BaTiO_3$ from Inframat Advanced Materials (Purity 99.95 %; Impurities: Ca, Fe , K, Mg, Na less than 0.001% and Sr less than 0.015%); $SrTiO_3$ from Alfa Aesar (Impurities: Silicon <165 ppm, Aluminium < 165 ppm, Barium < 0.8%, Carbon < 0.37 %, Iron <165 ppm, Chlorides < 35 ppm, Lead < 165 ppm) and from Inframat Advanced Materials (Purity: 99.95%. Impurities: Ca < 0.008 %, Fe < 0.006 %, K < 0.005 %, Mg <0.005 %, Na <0.006%).
Single crystals: $SrTiO_3$: <100> oriented; from MaTecK; Dimensions of the plate: 4.98x4.85 $mm^2$, thickness =0.51 mm.; $BaTiO_3$: <100> oriented; from GB Group, Crystals Land; Dimensions of the plate: 10x10 $mm^2$, 0.5 mm thick; $(Ba_{0.975}Sr_{0.025})TiO_3$: [001] oriented; from FEE GmbH Germany; 2.5x3.5 $mm^2$, t=0.49 mm. Profilometer was used to measure the curvature of the sintered ceramics disks. Only roughness of the surface could be detected.

**S1 Details on electro-mechanical and pyroelectric measurements.** Schematic of the home-made dynamic press used for the piezoelectric charge-force measurements is given in Fig. S1. Details are described in Ref. 1 of Supplementary Information (SI). The measurement accuracy is mostly affected by the position of the sample in the sample holder. The precision is affected by the electrical noise, ground loops, charge drift due to a finite insulation resistance, and temperature variations (see also Methods). A small piezoelectric-like signal, up to about two orders of magnitudes smaller than in $(Ba_{0.60}Sr_{0.40})TiO_3$ (BST60/40), was observed with direct electro-mechanical measurements in $SrTiO_3$ crystal and ceramics. This signal is close to the noise level, and is similar to that captured when a piece of glass or sapphire single crystal were measured (see also Ref. 2 of SI, where a large signal was observed in $BaTiO_3$ and a very small in mica). There are several possible sources of this weak signal, including the following: (i) it could be related to an electronic pick-up from the driving signal that could not be eliminated; (ii) one cannot entirely exclude other origins, proper to the material. Some authors have reported pressure induced potentials that are related to inhomogeneous deformation of crystal lattice on the surface, and pressure modified electro-chemical potentials or space charges at grain boundaries. See Refs. 2-5 of SI. Grain boundaries as the sole origin of this small signal can be excluded because we observe such signal in some single crystals and glasses. It is important to have in mind that in ferroelectric and paraelectric materials described here, where electro-mechanical coupling is strong and its phase depends on sample orientation, we also observe pyroelectric signal. In those compositions in which electro-mechanical signals is on the level of the noise, no pyroelectric coupling is seen. Absence of pyroelectric signal in those samples could, however, be a question of the resolution of our instruments.

The dynamic pyroelectric measurements were made using the set-up described in Ref. 6 of SI.

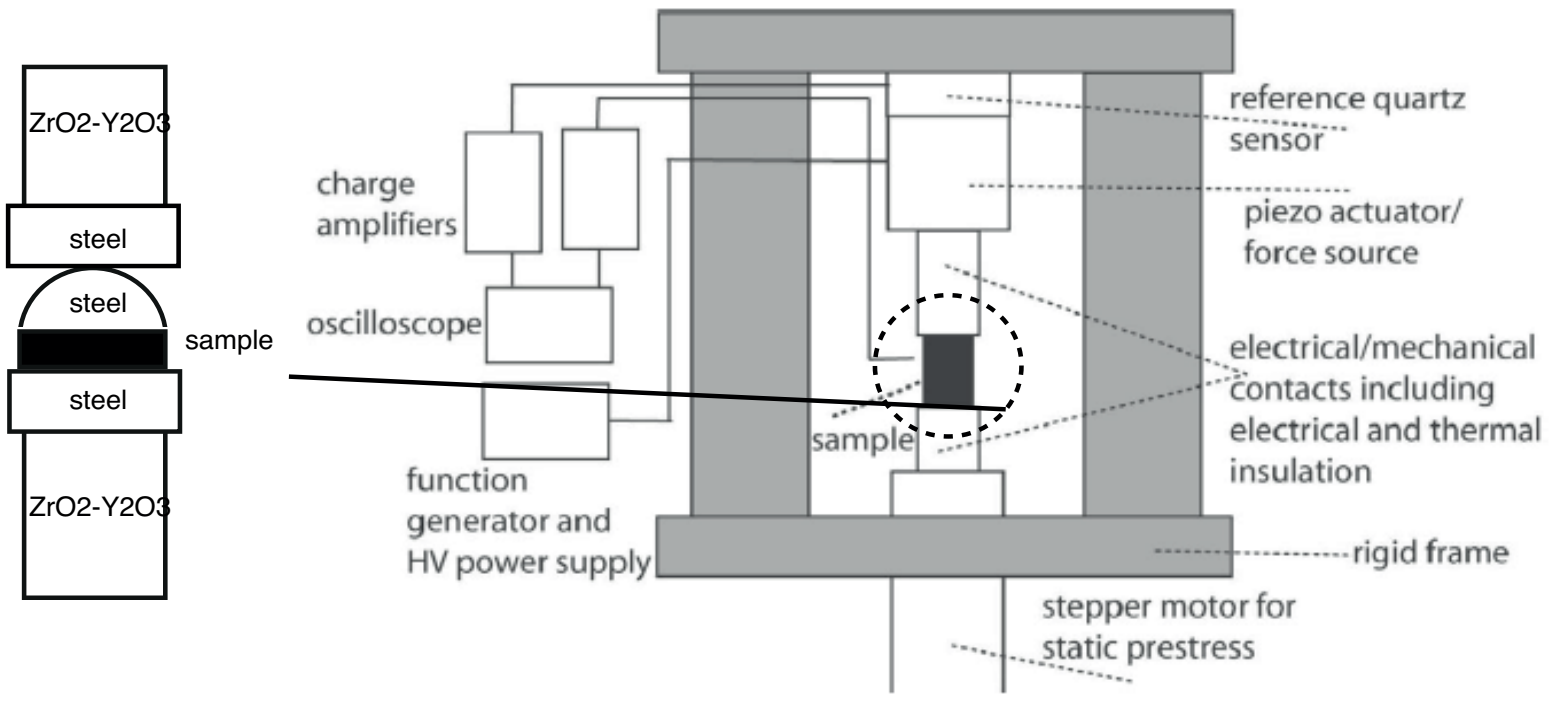

**Figure S1.1 The schematic of the dynamic press used for electro-mechanical measurements.** See Ref. 1 of SI for details.

**S2 Geometrical impact and built-in strain.** Let's consider possible flexoelectric origins of the charge due to geometrical or internal strain gradient when the stress is applied perpendicular to the symmetry axis of the trapezoid. We first discuss the trivial case of geometrical imperfections of the sample. The thickness of the sample shown in Fig. 1 of the article is by 0.17±0.01 mm wider (less than 9%) near the smaller than near the larger base of the trapezoid. To compensate for such imperfections and improve homogeneity of the pressure we applied the force via a hemispherical steel piece placed on one side of the sample (see Fig. S1.1). The small difference in the area of the two large sides of the sample is many times smaller than the difference in the area of the two bases of the trapezoid so that the charge cannot be explained through the geometrically generated strain gradient. In addition, the experiment was repeated on other samples of $(Ba,Sr)TiO_3$ solid solution with nominally parallel faces: disks, parallelepipeds – none of which should ideally exhibit significant strain gradient of geometrical origin – and all exhibited a comparable charge (within an order of magnitude). Even if the geometrical strain gradient were responsible for the charge, the corresponding flexoelectric coefficient would have to be larger than the one observed when stress is applied along the symmetry axis where the strain gradient is the largest.

Another possibility is that the sample exhibits internal strain gradient(s) originating from the preparation procedure or some nonobvious inhomogeneous stress distribution; the response could then be flexoelectric-like in nature. The magnitude of modulation of such hypothetical strain gradient by the external stress can be estimated. Assuming that the experimentally measured signal originates entirely from the intrinsic, lattice flexoelectric effect, taking the theoretical value of the $\mu$ coefficient (10 nC/m) and $P \approx 2 \times 10^{-6}$ C/m$^2$ (see Fig. 1), the relationship $P = \mu \partial S / \partial x$ leads to a strain gradient on the order of 200 m$^{-1}$ and equivalent stress gradient ≈30 TPa/m, obviously nonphysical values for a thick sample (see Ref. 2 of SI). Taking for this estimate the experimental value of $\mu$ would raise back the question of its large value and origin of the built-in strain gradient that would have to be as large as the shape-induced strain gradient.

**S3 Pyroelectric and piezoelectric responses above $T_C$ in BST60/40 ceramics and $BaTiO_3$ single crystals.** The Curie temperature $T_C$ (about 294 K during cooling and 307 K on heating) of BST67/33 lies close to the ambient temperature at which measurements were made. It may be argued that so close to the structural instability a ferroelectric state can be induced by heating and cooling of the sample by handling it during the experiment or by the static pre-stress (see Ref. 7 of SI) leading to a small piezoelectric response of the sample. This possibility, however, can be discarded because we see similar behavior in BST compositions for which $T_C$ is well below the ambient temperature, samples which have never been cooled to their ferroelectric phase and in samples that are heated dozens of degrees above their $T_C$.

Importantly, the pyroelectric and piezoelectric signals are seen in the samples that have never been in their ferroelectric phase (e.g., BST 60/40). This excludes a possibility that the pyroelectric and piezoelectric responses in the paraelectric phase are due to a "residual" ferroelectric phase or polarization trapped by defects (see Ref. 8 of SI).

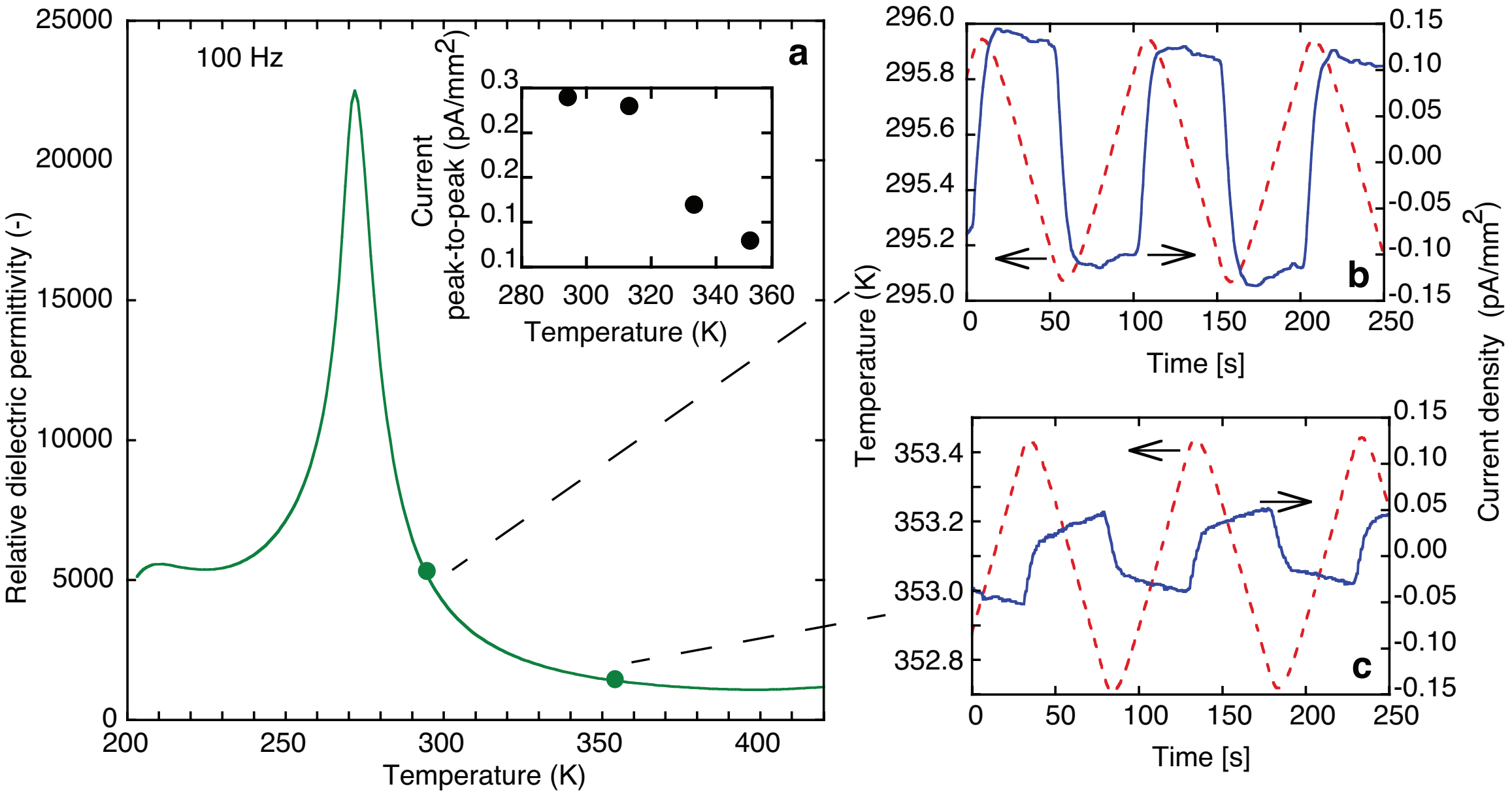

**Figure S3.1. Properties of BST60/40 above $T_C$.** The dielectric permittivity as a function of temperature (**a**) and pyroelectric response of BST60/40 ceramics at 294 K (**b**) and 353 K (**c**). The Curie temperature of this material is ≈272 K. Inset shows the amplitude of the pyroelectric current at four temperatures above the $T_C$.

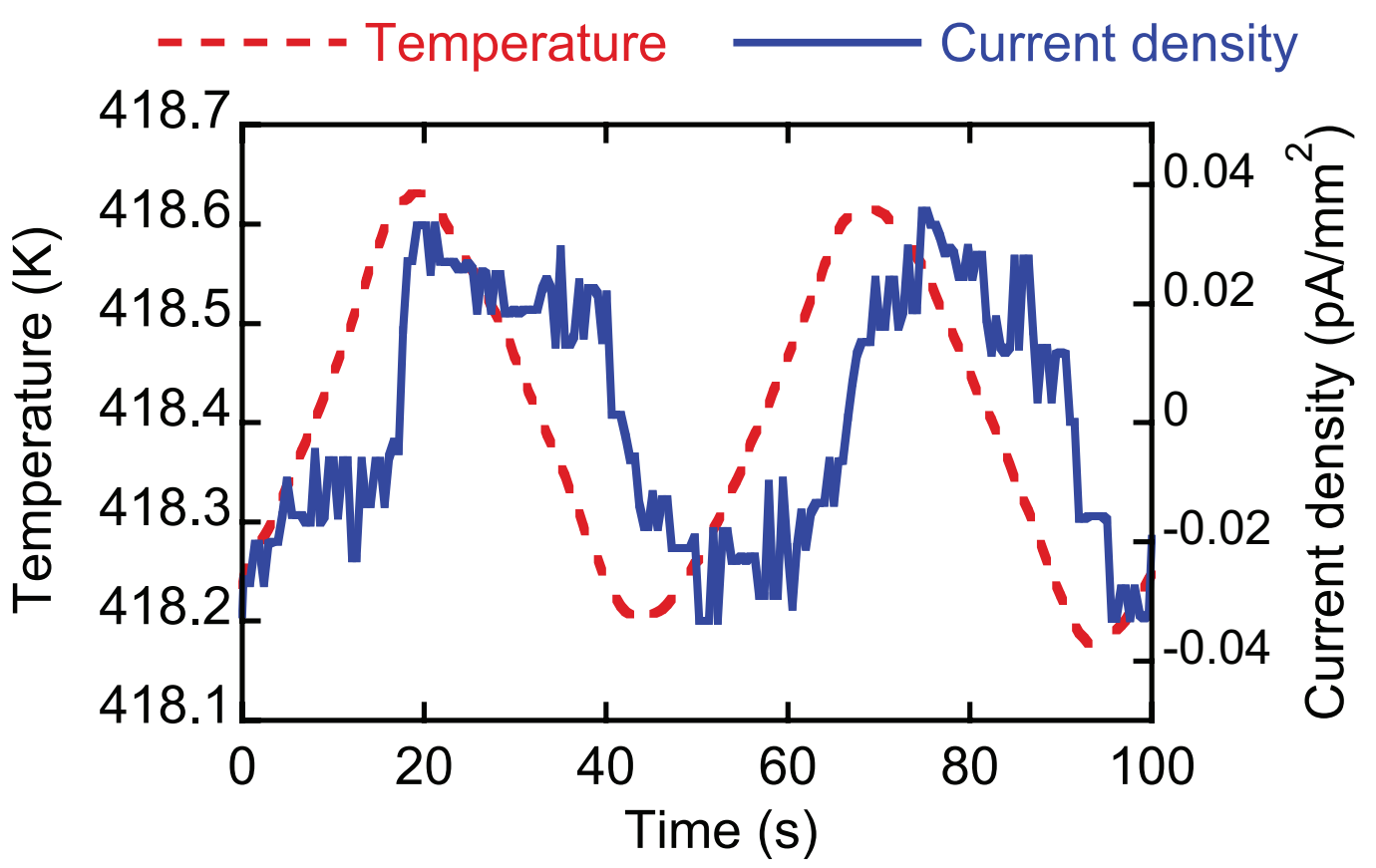

**Figure S3.2. The pyroelectric current in a $BaTiO_3$ single crystal above $T_C$.** The Curie temperature of this sample is 408 K and the measurements were made at 418 K. The noise in the current arises from the interference of temperature fluctuations in the ambient temperature (418 K) with the temperature modulation (< 0.5 K) used for measurements.

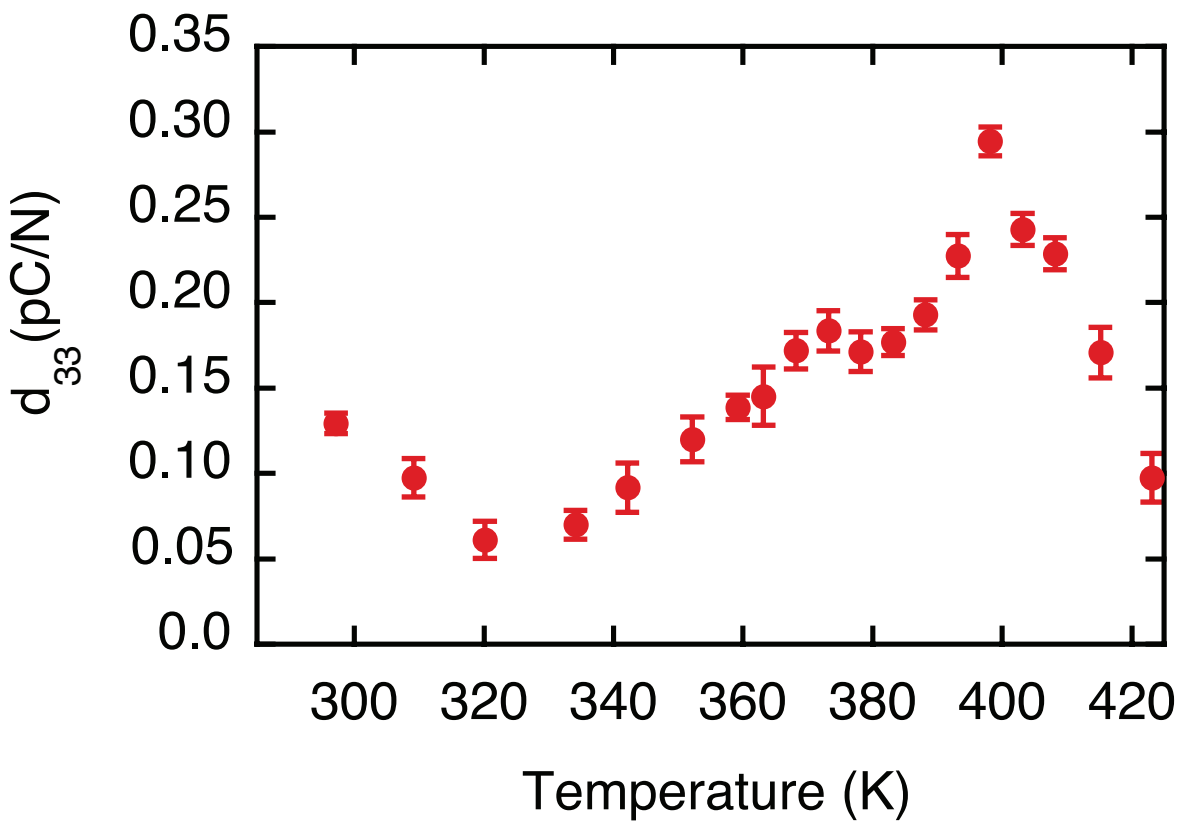

**Figure S3.3. The piezoelectric-like response in unpoled $BaTiO_3$ ceramic measured from 293 K to 420 K**. The Curie temperature of this sample is ≈398 K. Note that the magnitude of the response is similar above and below $T_C$, suggesting that in unpoled ferroelectric ceramics the symmetry breaking is not dominated by the imperfect cancelation of response of individual domains.

**S4 Consideration of the surfaces as the origin of the symmetry breaking.** Surface effects have been considered in the early literature on ferroelectrics as a possible origin of the polar character above $T_C$.[5] Here we give evidence that surface–related phenomena are not the dominant origin of the built-in polarization in samples investigated here. For pyroelectric measurements we compared electroded as-sintered samples and samples polished to different degrees of the surface roughness. For piezoelectric measurements we compared electroded and non-electroded samples, polished and as-sintered. The electrodes for most measurements were sputtered Au but we also tested samples sputtered with Au/Cr and Pt and samples painted with silver paste. Electro-mechanical and pyroelectric responses of the samples of a given compositions did not depend significantly on the electrode type. The size of the response varies in an irregular way with the state of the surface (polished or not) but the current direction is not dependent on it. A removal of about half of the thickness of an as-sintered sample (approximately 200 μm on each side of the sample) does not affect magnitude of the response significantly nor its direction. We thus conclude that the surface of a sample is not a dominant factor determining the direction or the magnitude of the built-in polarization.

**S5 Direction of the built-in polarization and position of the sample in the furnace.** To test effects of the samples' surrounding during preparation on direction of the built-in polarization, we carried out different tests. Most of the sintering runs were made in a box furnace with heating elements on the sides of the furnace. An $Al_2O_3$-based plate was placed on the bottom of the furnace chamber and different supports were placed between the alumina plate and the samples, such as Pt foil and Pt wires (to minimize contact area between the sample and Pt). Pt was needed to prevent reaction of the sample with the alumina plate. Most samples were exposed to air during the sintering. Some samples were placed in small alumina containers and completely buried in the powder of the same composition as the sample. Other samples were enveloped completely in a Pt foil during the sintering. Some samples were suspended in air by Pt wires during sintering. Some samples were sintered in a tube furnace, to verify effects of distribution of heating elements on the direction of polarization within samples. Some disk-shaped samples were placed vertically with respect to the bottom of the box furnace. In each case we observed built-in polarization with the definite direction with respect to the orientation of the sample in the furnace. The pyroelectric and piezoelectric responses of the samples produced by these different methods vary within an order of magnitude, providing that parameters such as sintering time and temperature were kept similar. Importantly, we observed symmetry breaking in BST60/40 samples sintered by spark plasma sintering and provided by Dr. P.-E. Janolin from Ecole Central Paris. In those samples the pyroelectric response is more than an order of magnitude smaller the in ours but those samples exhibited strongly suppressed dielectric anomaly, and were dark in color suggesting a large concentration of unidentified defects.

**S6 Space charges, electret behavior and thermally stimulated currents.**

The thermally stimulated currents (TSC) exhibit characteristic peaks only during the heating but not during the cooling of the samples, as shown in Fig. S6.1.

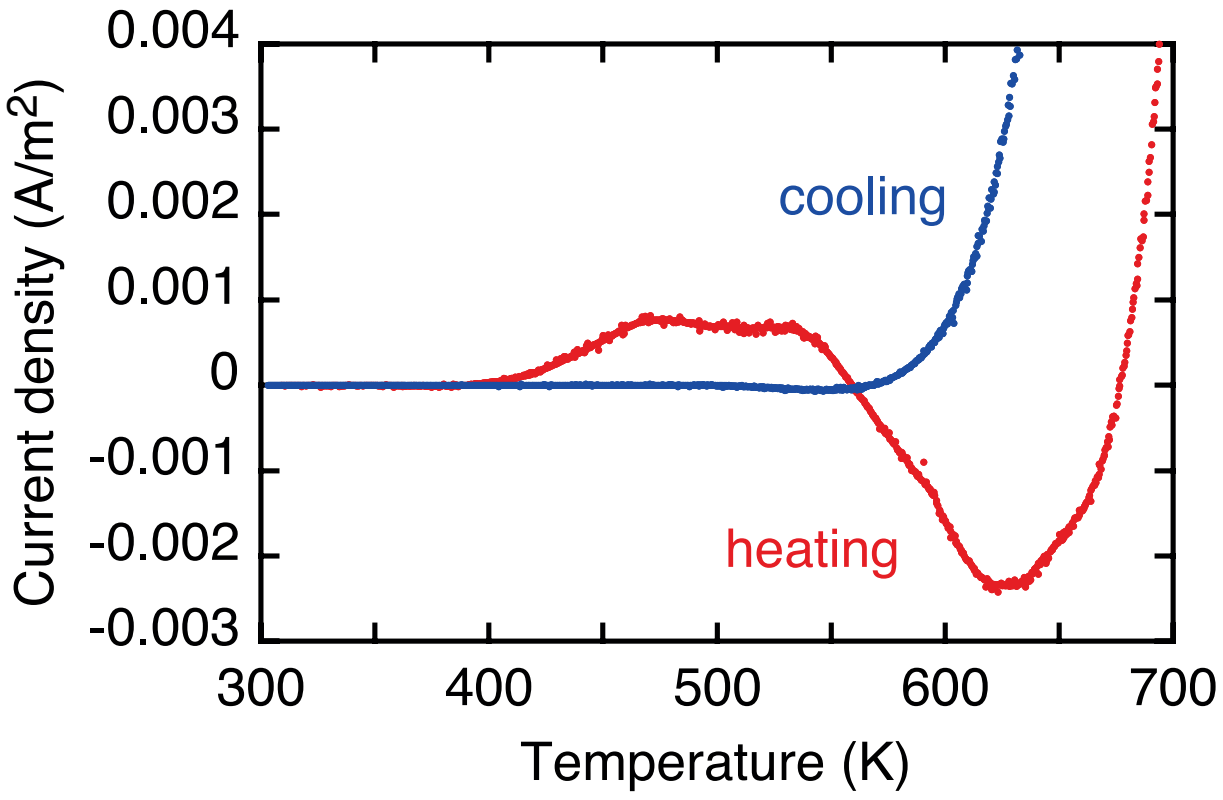

**Figure S6.1. The thermally stimulated currents measured during heating and cooling for the BST60/40 sample shown in Fig. 4a**. The absence of TSC peaks during cooling is an indication that the peaks observed on heating are due to thermally stimulated charge de-trapping and not a phase transition, as described by Roth *et al.* for relaxor ferroelectrics (see Ref. 9 of SI). The latter is expected to give a signal through a sudden release of charges also on cooling.

The presence of mobile space charges is demonstrated by poling (forming an electret) of $SrTiO_3$ and $(Ba_{0.6}Sr_{0.4})TiO_3$ ceramics. The samples were poled at room temperature applying 2500 V/cm for 10 minutes at 296 K. The $SrTiO_3$ sample, which does not show any pyroelectric current when examined in as-sintered state, exhibited pyroelectric current after poling. The pyroelectric current amplitude decayed relatively quickly and within ≈70 hours no signal could be observed in $SrTiO_3$. The signal also disappeared if the sample was heated to 323 K shortly after the poling. With respect to the as-sintered state, $(Ba_{0.6}Sr_{0.4})TiO_3$ sample exhibited enhanced signal when poled in the direction of the built-in polarization, while it showed reduced signal and of the opposite polarity when poled against the build-in polarization, Fig. S6.2. Within 2 days, however, the poling effect dissipated and the sample exhibited its usual built-in polarization attained after sintering.

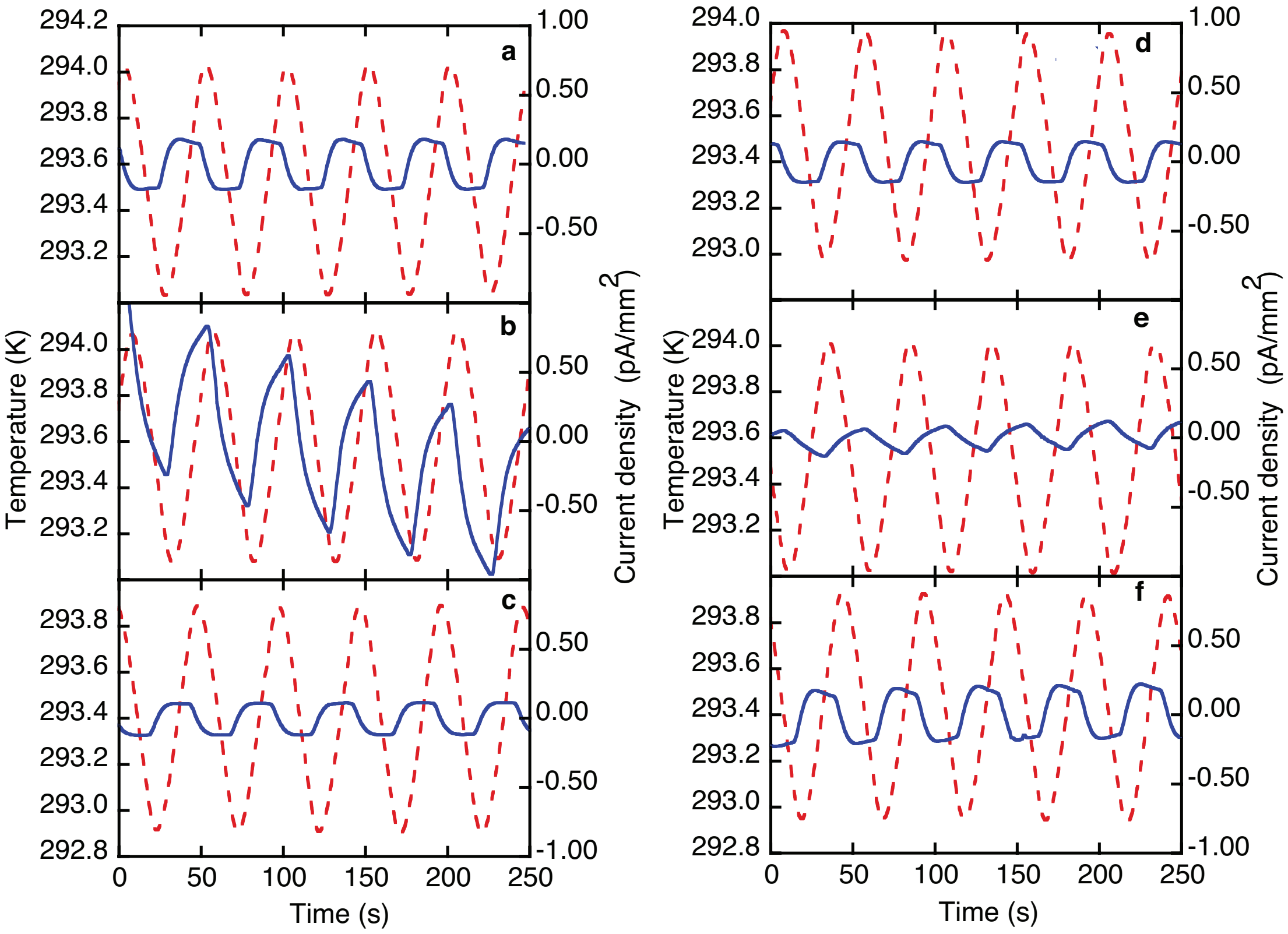

**Figure S6.2 Formation of an electret in BST60/40**. The pyroelectric response (solid line) and the driving temperature (dashed line) at ambient temperature. (**a,b,c**) show data for a sample poled in the direction of built-in polarization: (**a**) before the poling; (**b**) immediately after the poling; (**c**) two days after the poling, illustrating decay of the space charge. (**d,e,f**) show a sample poled opposite of the direction of the built-in polarization: (**d**) before the poling; (**e**) immediately after the poling; (**f**) two days after the poling, illustrating decay of the space charge. Note that the poling against the built-in polarization, shown in (**e**), changes the phase of the pyroelectric signal, indicating that in this case the field-induced and the built-in polarization have opposite directions and that the field-induced polarization dominates. However, after the decay of the space charge, the built-in polarization dominates and the polarization direction reverts to the original direction attained after the sintering.

Even though no pyroelectric signal is observed in as-sintered $SrTiO_3$ ceramic samples, the TSC currents were sensitive to the orientation of the samples, Fig. S6.3. This is consistent with the presence of a built-in polarization in ceramic $SrTiO_3$, but whose variation with temperature is probably too weak to be observed through measurements of pyroelectric currents at ambient temperature.

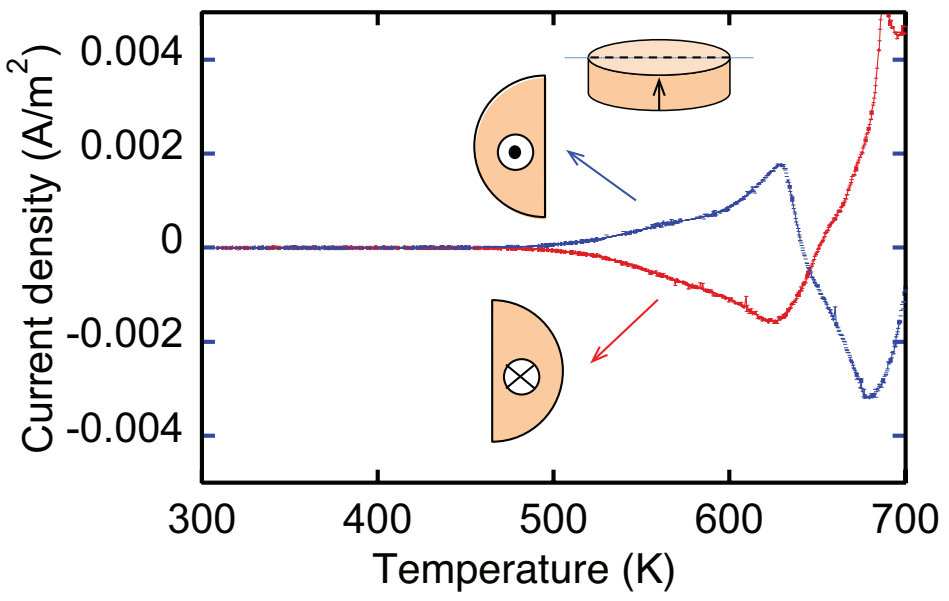

**Figure S6.3. The TSC in a $SrTiO_3$ ceramic during a heating run**. The sample was cut into two halves and TSC measured on both halves separately, one half being flipped over with respect to the other. The direction of the current depends on the orientation of the halves indicating sample's polar character.

## Supplementary References:

1. Barzegar, A., Damjanovic, D. & Setter, N. The effect of boundary conditions and sample aspect ratio on apparent $d_{33}$ piezoelectric coefficient determined by direct quasistatic method. *IEEE Trans. UFFC* **51**, 262-270 (2004).
2. Bursian, E. V. & Zaikovski, O. I. Changes in the curvaure of a ferroelectric film due to polarization. *Soviet Physics - Solid State* **10**, 1121 (1968).
3. Raj, R., Cologna, M. & Francis, J. S. C. Influence of externally imposed and internally generated electrical fields on grain growth, diffusional creep, sintering and related phenomena in ceramics. *J. Am. Ceram. Soc.* **94**, 1941-1965, doi:10.1111/j.1551-2916.2011.04652.x (2011).
4. Pannikkat, A. K. & Raj, R. Measurement of an electrical potential induced by normal stress applied to the interface of an ionic material at elevated temperatures. *Acta Mater.* **47**, 3423-3431 (1999).
5. Lines, M. E. & Glass, A. M. *Principles and Applications of Ferroelectrics and Related Materials*. (Clarendon, 1979).
6. Daglish, M. A dynamic method for determining the pyroelectric response of thin films. *Integrated Ferroelectrics* **22**, 473-488 (1998 ).
7. Picht, G. *et al.* Critical mechanical and electrical transition behavior of $BaTiO_3$: The observation of mechanical double loop behavior. *J. Appl. Phys.* **112**, 124101, doi:10.1063/1.4767059 (2012).
8. Darlington, C. N. W. & Cernik, R. J. "The ferroelectric phase transition in pure and lightly doped barium titanate," *J. Phys.: Condensed Matter,* **3**, 4555-4567 (1991).
9. Roth, M., Mojaev, E., Dul'kin, E., Gemeiner, P. & Dkhil, B. Phase transition at a nanometer scale detected by acoustic emission within the cubic phase $Pb(Zn_{1/3}Nb_{2/3})O_3$-x$PbTiO_3$ relaxor ferroelectrics. *Phys. Rev. Lett.* **98**, 265701 (2007).